\documentclass[a4paper]{article}

\usepackage{amsmath}
\usepackage{amsfonts}
\usepackage{amssymb}
\usepackage{amsthm}
\usepackage{xcolor}
\usepackage{algorithm}
\usepackage{algpseudocode}
\usepackage{graphicx}
\usepackage{enumitem}
\usepackage{hyperref}

\hypersetup{
    colorlinks=true,
    linkcolor=blue,
    filecolor=magenta,      
    urlcolor=cyan,
}

\providecommand{\algorithmname}{Protocol}
\floatname{algorithm}{\protect\algorithmname}

\newtheorem{proposition}{Proposition}
\newtheorem{observation}{Observation}

\newtheorem{definition}{Definition}

\newcommand{\nimrod}[1]{\textcolor{red}{Nimrod says: #1}}

\newcommand{\ouri}[1]{\textcolor{magenta}{Ouri says: #1}}

\newcommand{\hidden}[1]{}

\title{A Consensus Protocol for e-Democracy}
\author{Ouri Poupko \and Nimrod Talmon}

\begin{document}

\maketitle

\begin{abstract}
    Given that Proof-of-Work (PoW) and Proof-of-Stake (PoS) are plutocratic, and other common consensus protocols are mostly permission-based, we look for a consensus protocol that will suit the needs of e-Democracy. In particular, what we need is a distributed ledger that will record and, to the possible extent, execute the public will. We propose a combination of any given permission-based protocol together with a trust graph between the nodes, which supplies the required permission for new nodes. As a result, the consensus protocol reaches consensus at every iteration between a known list of agents and then updates this list between iterations. This paper is based on prior work that shows the conditions under which a community can grow while maintaining a bounded number of byzantines~\cite{RN367}. It combines a permission-based consensus protocol (such as pBFT~\cite{RN581}) with a community expansion algorithm (such as the one in the prior work) to arrive at a consensus protocol in which the set of agents can change in time, while being sybil-resilient.
\end{abstract}

\section{Introduction}

In e-Democracy operating via a distributed ledger, we would want the nodes to be the citizens; i.e., one person - one vote, all the way down to the consensus infrastructure. We start by noticing that in an e-Democracy based on a distributed ledger there are two types of required consensuses: \emph{mining consensus} and \emph{voting consensus}.

\paragraph{Mining consensus}
To maintain a distributed ledger all participating nodes need to be in consensus about what is written in the ledger. One key to achieve such consensus is the separation between what is claimed and the validity of the claim. In a slightly different view it is the difference between agreeing that Alice votes for Bob and agreeing that Bob is the chosen leader. We call the "what is claimed" consensus - the mining consensus. It is easier to agree on "what is claimed" (Alice votes for Bob), especially as cryptography gives Alice a way to prove that this is what she said, by signing this statement with her private key. If one node in the distributed ledger refuses to register her claim, she can easily turn to another node for the registration process.

The threat model for this consensus is the byzantine problem - the problem of confusion between the ranks. This is the problem where some of the \emph{honest} nodes believe that the majority of the nodes agree one way, and some other \emph{honest} nodes believe that the majority of the nodes agree another way. If Alice is malicious, this can lead to a double spend attack where Alice convinces some of the nodes that there is consensus (over majority of honest nodes) that she voted for Bob, and in the same time she convinces some other nodes that there is a consensus (over majority of honest nodes) that she voted for Carol. This threat model is completely resolved (using pBFT~\cite{RN581} for example) if the number of participants is at least $n=3f+1$, where $f$ is the number of byzantines.

\paragraph{Voting consensus}
The consensus over the democratic validity of the claim (or agreeing that Bob is the chosen leader) is what we call the voting consensus. In some social manner, it is harder to achieve, since people tend to disagree even on objective facts when it compromises their emotional believes. However, given an agreed voting mechanism (including a reliable registrar) and given an agreed list of voters, it is easy to reach a democratic decision simply by counting majority of votes.

The threat model for this consensus is when the corrupted voters infiltrate enough Sybil voters in order to overcome majority. In this attack everyone will reach consensus that Carol is the leader, even though majority of genuine identities (honest and corrupted) voted for Bob. The way to overcome this threat, given a known bound on the ratio of Sybils, is to require super-majority of votes. This method looses liveness (that is, the community is stuck in the current state of affairs) when $n<3f+1$, however this is a linear degradation. If $n=3f+1$, then a super-majority of $2f+1$ should be required, meaning that even a single genuine identity (together with $f$ sybils) can block any proposal.

\subsection{Reaching consensus for e-Democracy}\label{subsec: cons prot}

To reach voting consensus we need to have a trusted list of voters. To have a trusted list of voters we need a trusted registration system with mining consensus. It is not clear however how to achieve mining consensus without having voting consensus over who is a valid miner. This is the bootstrap problem. One way to solve it is by trusting a group of founders to build a small community and grow the community step by step. This however gives a great deal of power to the founders in moulding the shape of the community. Our approach in general is to distribute the decision making power of the group among all members of the group, in a way that each new member, once joined, will have exactly his proportional share in the decisions of the group, including in processes that started before he joined. Specifically in this paper we deal with two basic processes, 1 - deciding who is in the group, and 2 - logging the group history. These two functions merge on the specific task of logging the history of who is in the group (who joins when).

There is a plethora of academic and industry work about consensus protocols. Largely, they are divided into permissionless protocols, like POW and POS, where anyone can freely join the group of agents running the protocol, and permissioned protocols, where all the agents maintain a list of all participating agents and the protocol reach consensus explicitly among this group. Known permissionless protocols lean on resources burning or owning that make them less ideal for democratic governance. POW has a negative impact on the environment and POS is plutocratic. Permissioned protocols are simpler, but they are not distributed enough. This paper describes a consensus protocol that is semi-permissioned. It starts with a permissioned protocol, but distribute the permission authority among all agents. This way it maintains the simplicity of permissioned protocols, while gaining the scalability of permissionless protocols.

\subsection{Related Work}

We refer to Proof-of-Work (POW) and to bitcoin~\cite{RN346} as its first outstanding implementation, as a starting point to our discussion. proof of work enables the creation of a distributed ledger which is permissionless. Anyone can join the network that is running the distributed ledger, without affecting the safety of the protocol, or even improving the safety, as it grows as the network is further distributed. However it holds grave consequences. In order to achieve safety in a permissionless group, the participants keep competing among themselves in solving the small hash problem, all for the sake of creating an order between them. A computation whose output is completely useless once the order is decided. As a result, the bitcoin network of processing units consumes more power than an average country~\cite{RN233}. The second problem is that the POW protocol is plutocratic by nature. Richer participants can invest more in hardware, which makes them even more stronger and richer participants.

Proof-of-Stake (POS)~\cite{RN449} can resolve the electricity power arm racing problem. the order of participation can be determined by the amount of stake the participant has in the network. It is also permissionless. Anyone can join and as more participants join the system gets more robust. There is no need for solving useless problems. Yet, again, by design, POS is plutocratic as well.

On the other end of the scale, pBFT~\cite{RN581} is considered the flag protocol for permissioned distributed ledger. If the group of participants is known, two rounds of verification messages among the group are enough to reach consensus. As the number of messages for each consensus round is of order $O(n^2)$, this protocol does not scale well for more then several hundreds, or at most thousands of participants.

There are multiple attempts to create distributed ledger protocols that let anyone join on their own act. Byzcoin~\cite{RN582} uses POW as a Proof-of-Membership, and then uses pBFT for a more efficient consensus among the members. Need to elaborate further.

The Proof-of-Identity concept relies on some off-chain mechanism to identify the participants, which can then prove their identity with cryptographic means to participate in the distributed ledger network. Though the concept is widely discussed, we did not encounter yet a distributed ledger implementation using it.

In Proof-of-Personhood~\cite{RN353} Borge et al. propose to conduct pseudonym parties, where group of people physically meet in one place to create cryptographic tokens for the participants. Though they discuss scalability briefly by proposing to conduct multiple parties simultaneously in different regions, it is hard to see such parties scale to worldwide proportions.

Two versions of Ethereum, parity and geth, implement the aura and clique consensus protocols that use Proof-of-Authority. A Proof-of-Authority protocol~\cite{RN355} uses a chosen set of miners which in turn put their reputation as a stake.

The Stellar protocol, Federated byzantine agreement~\cite{RN583}, runs the consensus protocol within quorum slices. Each node marks which quorum slices it trusts (and hence part of). The protocol ensures that consensus is achieved between quorums with intersection, but it does not guarantee global consensus.

In 1997, Rubinstein and Kasher propose \cite{RN572} to look on the question of who is a member of a given society as a social choice question, aggregating the opinions of the members of the society themselves. It is in the spirit of this approach that we proposed \cite{RN367} a method for a sybil resilient community growth.

\subsection{Paper structure}

We present our protocol in stages, analyzing each stage and its validity. In section \ref{section: grow} we start with a simple protocol that only adds new nodes in each iteration. In section \ref{section: remove edges} we handle the case of nodes leaving the trust graph. We finalize with a complete protocol in section \ref{section: join}, that not only adds and removes nodes, but also carefully merges history between joined communities.

\section{Preliminaries}

We put forward our two main tools, namely a consensus protocol and a function that maintains a sybil-resilient community.

\subsection{Consensus Protocl}

Among permissioned protocols, pBFT is probably the most common reference solution. We take it as an example for the combined protocol proposed inhere, but generalize it as follows. A permissioned consensus protocol in the context of this paper has the following characteristics.

\begin{definition}[\texttt{Consensus}]\label{def: consensus}
A \emph{\texttt{Consensus}} function is a protocol that adheres to the following requirements:
    \begin{enumerate}
        \item It is a replicated state machine, where each node maintains an internal state and a set of accepted events. The state machine is deterministic, and all nodes start from the same initial state, therefore if all nodes agree on the order of incoming events, then all nodes maintain the same state all the time.
        
        \item It is asynchronic in the sense that the network connecting the nodes may fail to deliver messages, delay them, duplicate them or deliver them out of order. We do assume one relaxation to this requirement, that by some unknown bounded time a message delivered from a proper node to another proper node will eventually arrive.
        
        \item It maintains safety, that is it satisfies linearizability between operations, similar to a centralized implementation.
        
        \item It maintains this list of requirements under the assumption that less than $\beta_{cp}\cdot n$ nodes are faulty.\label{req: bounded byzantines}
    \end{enumerate}
\end{definition}

Formally, the \texttt{Consensus} function returns either \texttt{true} or \texttt{false}, whether all nodes accept the given event as the next event. This is a slight adaptation to the more common implementation of a replicated state machine, where an event stays in queue until it is accepted by the consensus protocol. We do it for the clarity of the protocols to follow.

\subsection{Resilient Community Function}

In a prior work \cite{RN367} we identify the conditions under which a community can grow while maintaining a bound on the number of sybils. For the purpose of this work we generalize the following community growth function:

\begin{definition}[\texttt{ResilientCommunity}]\label{def: resilient community}
A \emph{\texttt{ResilientCommunity}} function is a protocol that adhere to the following requirements:
    \begin{enumerate}\label{thm: comm growth}
        \item It accepts as input a graph $G=(A,V,E)$ where $V=H\uplus C\uplus S$ (honest, corrupt and sybils respectively) and $A\subset V$ is a set of nodes with bounded number of corrupt and sybils.
        \item It accepts as input an additional set of nodes $A'\subset V$.
        \item It assumes a bounded number of corrupt nodes in $A$: $\frac{|A\cap C|}{|A|}\leq\gamma$.
        \item It assumes a bounded number of byzantine in $A$: $\frac{|(A\cap C)\cup(A\cap S)|}{|A|}\leq\frac{1}{2}$.
        \item it outputs true if $\frac{|(A\cup A')\cap (C\cup S)|}{|(A\cup A')|}\leq\beta_{rc}$.
    
    \end{enumerate}
\end{definition}

In plain words, the above function examine the graph as the users add nodes to the community $A$. The function accepts the new set of nodes, if it can guarantee that the number of byzantine in the community stays below $\beta_{rc}$. Our community growth method~\cite{RN367} achieves that by inspecting the connectivity of the graph. We also assume, though it is not articulated above, that there is always a way to add new nodes to the graph under the above conditions (liveness), so a community that maintains a bound on the number of corrupt nodes can continue to grow iteratively.

\section{From Community Expansion to Consensus}\label{section: grow}

The protocol in this section describes the handling of a new event sent to the replicated state machine. It starts with running a permissioned consensus protocol that adheres to the requirements in definition \ref{def: consensus}. This consensus protocol uses the list of nodes of the graph that the replicated state machine stores in its current state as the list of agents that should reach consensus. Once the event is accepted, the protocol updates the graph structure according to the community growth procedure described in definition \ref{def: resilient community}.

It is custom in distributed ledgers, like bitcoin, to differentiate between the computer nodes that participate in the consensus protocol (the miners) and the users that sign the transactions to be logged in the ledger (the clients). In the approach given here there is one pool of identities. Both the nodes that participate in the consensus protocol, and the users that sign the transactions to add edges to the graph, are the same identities. When an edge is created between $v_1$ and $v_2$, it is $v_1$ and $v_2$ that sign the transaction, and it is the same $v_1$ and $v_2$ that will participate in the consensus protocol, once they enter the community $A$.

\subsection{The protocol}
Note that the weakness of the \texttt{ResilientCommunity} protocol is that it depends on an external assumption that the number of corrupt nodes within the community $A$ is bounded. We capture this external assumption by the following oracle definition.

\begin{definition}[\texttt{CorruptionOracle}]
A \emph{\texttt{CorruptionOracle}} is a magic function that verifies that the corrupt nodes portion in a community $A$ is less than $\gamma$.
\end{definition}

\begin{algorithm}[t]
\caption{Semi permissioned consensus protocol -- event handling}\label{alg: protocol}
\begin{algorithmic}[1]
\State init $state$ with an empty graph
\State init $logger$ with an empty log
\Function{HandleEvent}{$state$, $event$}
    \State $G\gets$ \Call{$state$.getGraph()}{}
    \State $(A,V,E)\gets$ \Call{$G$.deconstruct()}{}
    \If{\Call{IsValid}{$event$}}
        \If{\Call{Consensus}{$A$, $event$}}
            \State\Call{$logger$.log}{$event$}
            \If{\Call{$\mathit{event}$.getType()}{} is CONNECT\_IN\_G}
                \State $\mathit{edges}\gets$\Call{$\mathit{event}$.getEdges()}{}
                \ForAll{$\mathit{edge} \in \mathit{edges}$}
                    \State $E\gets E\cup \{\mathit{edge}\}$
                \EndFor
            \EndIf
            \If{\Call{$\mathit{event}$.getType()}{} is EXTEND\_COMMUNITY}
                \State $A'\gets$\Call{$\mathit{event}$.getCommunity()}{}
                \State $C_1 \gets $\Call{CorruptionOracle}{$A \cup A'$}
                \State $C_2 \gets $\Call{ResilientCommunity}{$G$, $A'$}\label{line: resilient community}
                \If{$C_1$ and $C_2$}
                    \State$A\gets A\cup A'$
                \EndIf
            \EndIf
        \EndIf
    \EndIf
\EndFunction
\end{algorithmic}
\end{algorithm}

The protocol starts with an empty graph and log. This guarantees that all replicas start with the same initial state. When the consensus community $A$ is empty we assume that the $\mathit{Consensus}$ function will accept any event it is given (if no one objects, then everyone agree). We discuss the bootstrapping of the protocol further in subsection \ref{subsection: joining networks}. The first thing the protocol does is to extract the graph structure from its internal state (lines 4,5). Once the validity of the input event is checked, the protocol runs the consensus protocol, and if successful (all replicas accept the event), it is written to the log (added to the ledger). The protocol then reacts to two types of special events. The first (lines 9-14) simply adds edges (and possibly new nodes that are connected to them) to the underlying graph $G$. The second (lines 15-22) is a request to grow the consensus community $A$. It is accepted under two conditions, that the oracle verifies that the bound on corrupt identities is not crossed, and that the resilient community method verifies that the bound on byzantines was not exceeded. The protocol is then ready to handle the next event.

\subsection{Validity propositions}

Under the assumption that honest nodes in the generation of the community graph (nodes without edges to sybil nodes) will behave honestly also in the consensus protocol (will not be faulty), as long as the amount of corrupt nodes is bounded, protocol \ref{alg: protocol} is a valid consensus protocol. More formally:

\begin{proposition}[Honests are honest]\label{prop: h are h}
Let $\gamma$ be the required bound on corrupt nodes in definition \ref{def: resilient community} to achieve a bound $\beta_{rc}$ on the ratio of byzantine nodes. Assume:
    \begin{enumerate}
        \item $\beta_{rc}\le\beta_{cp}$.\label{assumption: h are h b less b}
        \item A node $v\in H$ is not faulty in the consensus protocol.\label{assumption: h are h h non faulty}
        \item \emph{\texttt{CorruptionOracle}} ensures less than $\gamma$ corrupt nodes in every iteration of protocol \ref{alg: protocol}.\label{assumption: h are h co less g}
    \end{enumerate}
    Then Protocol \ref{alg: protocol} maintains the requirements of definition \ref{def: consensus}.
\end{proposition}

\begin{proof}
The call to \texttt{ResilientCommunity} in line \ref{line: resilient community}, together with assumption \ref{assumption: h are h co less g}, assure that the ratio of byzantine nodes in the graph is not more than $\beta_{rc}$. Assumption \ref{assumption: h are h b less b} assures that the ratio of byzantine nodes is also not more than $\beta_{cp}$. It follows from assumption \ref{assumption: h are h h non faulty} that requirement \ref{req: bounded byzantines} in definition \ref{def: consensus} holds. All other requirements are indifferent to protocol \ref{alg: protocol}.
\end{proof}

In the sybil resilient community growth paper~\cite{RN367} we showed that using vertex expansion a community can keep the byzantine ratio below $\beta$ as long as the vertex expansion of the graph is higher than $\frac{\gamma}{\beta}$. We also showed that a graph with vertex expansion $\Phi_v\ge\frac{2}{5}$ is quite feasible. It follows from proposition \ref{prop: h are h} that if the community can make sure, that in any given point in time the ratio of corrupt identities in the community is less than $\frac{2}{15}$, then it can maintain a community with less than $\frac{3}{15}$ sybil identities, and over all less than $\frac{1}{3}$ byzantines. Under these conditions such a community can safely use protocol \ref{alg: protocol} as a consensus protocol within the community.

The assumption that honests are honest marks one end of the spectrum. At the other end we can make an opposite assumption that there is no intersection between byzantines-to-the-consensus-protocol and byzantines-to-the-resilient-community-protocol. Under this assumption protocol \ref{alg: protocol} can tolerate up to $\beta_{cp}$ byzantines of the first type (anarchists) and $\beta_{rc}$ byzantines of the second type (corrupt and sybils).

\begin{proposition}[Anarchists are not corrupt]
Let $\gamma$ be the required bound on corrupt nodes in definition \ref{def: resilient community} to achieve a bound $\beta_{rc}$ on the ratio of byzantine nodes. Assume:
    \begin{enumerate}
        \item \emph{\texttt{CorruptionOracle}} ensures less than $\gamma$ corrupt nodes in every iteration of protocol \ref{alg: protocol}.\label{assumption: a are nc co less g}
        \item An additional procedure \emph{\texttt{AnarchistOracle}} (not listed in protocol \ref{alg: protocol}) ensures less than $\beta_{cp}$ faulty nodes in every iteration of protocol \ref{alg: protocol}.\label{assumption: a are nc fa less b}
    \end{enumerate}
    Then Protocol \ref{alg: protocol} maintains the requirements of definition \ref{def: consensus}.
\end{proposition}

\begin{proof}
As long as some external protocol ensures that less then $\beta_{cp}$ nodes are faulty, the \texttt{Consensus} procedure maintains its requirements, regardless of the underlying graph between the nodes. Protocol \ref{alg: protocol} therefore maintains the same requirements as well.
\end{proof}

\subsection{Growing the Network as the Graph Grows}\label{subsection: joining networks}
To better understand the interplay between nodes of the consensus protocol and the clients that commit transactions, consider first the code for the main loop of the protocol.

\begin{algorithm}
\caption{Semi permissioned consensus protocol -- main loop}\label{alg: protocol main}
\begin{algorithmic}[1]
\Function{MainLoop}{$neighbour$}
    \If{$neighbour$ is not empty}
        \State $history\gets$\Call{$neighbour$.getLog()}{}
        \ForAll{$event\in history$}
            \State\Call{HandleEvent}{$state$,$event$}
        \EndFor
    \EndIf
    \While{$true$}
        \State $event\gets$ wait for event
        \State\Call{HandleEvent}{$state$,$event$}
    \EndWhile
\EndFunction
\end{algorithmic}
\end{algorithm}

To demonstrate the interplay further, consider the following inefficient, yet explanatory example.

\begin{enumerate}[itemindent=2.5em,label={\bfseries Step \arabic*:}]
\item The first computer $A$ starts running. It initializes with an empty graph.
\item $A$ receives an event to add $B$ and $C$ to its underlying graph $G$.
\item $A$ receives an event to add $B$ and $C$ to the community of trusted nodes.
\end{enumerate}

Note that at this point we expect that the \texttt{Consensus} procedure, when running the consensus protocol on an empty set of nodes, will return true (if no one objects, then everyone agrees). Therefore the first two events are consensually accepted. Secondly, at this point node $A$ is blocked from receiving any further events, as such event will only be accepted by consensus among the set $\{B,C\}$, but nodes $B$ and $C$ are not yet in the network. From $A$'s perspective, both $B$ and $C$ are byzantine at this point.

\begin{enumerate}[resume*]
\item Computers $B$ and $C$ start running, initialized with an empty graph.
\item Computers $B$ and $C$ join the network of $A$ by querying it for the history of events. $B$ and $C$ then run the history of events locally and reach the same internal state as $A$.
\end{enumerate}
From this point all three nodes are synchronized and can run the consensus protocol together to continue accepting new events. Note that $A$ can participate in this process, though all three nodes, when aiming for consensus, adhere only to the messages from $B$ and $C$, as $A$ is not part of the trust graph.

\begin{figure}[t]
\centering
\includegraphics[width=8.5cm]{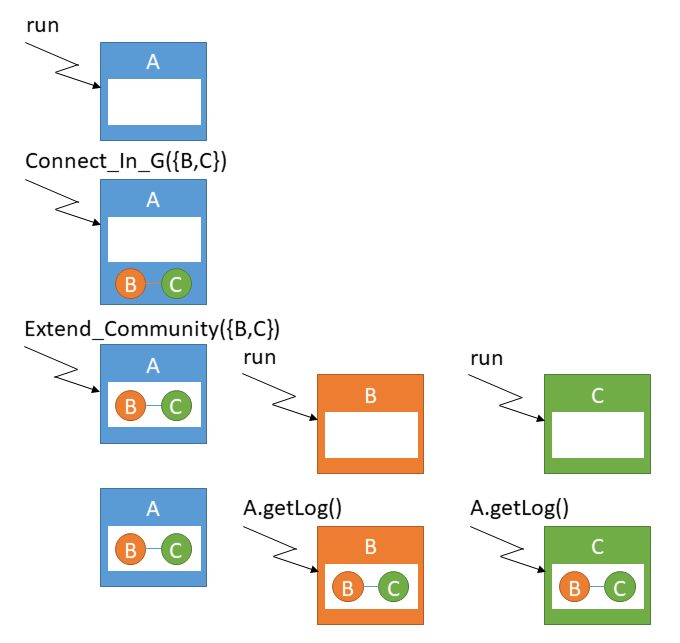}
\caption{Interplay between computer nodes in the network and identity nodes in the graph. Each rectangle is a computer. The graph within represents its internal state.}
\label{figure: inteplay}
\end{figure}

\section{Edge Removals}\label{section: remove edges}

Removing an edge from a trust graph in the above setting is problematic. While the community can add nodes one at a time while maintaining graph connectivity, when removing an edge the graph might split in the middle, and it is not clear which are the nodes that should be removed. A better solution for community growth, with the ability also to shrink back, should probably include not only the possibility to create edges between nodes that wish to be in the community, but also marking nodes as 'bad' nodes that should not appear in the graph. One way to do it might be through creating edges with negative weights, to capture that node $v_1$ states that node $v_2$ should not be in the community. This will require directional graphs, as one can expect pairs of nodes to agree simultaneously to be part of the same community, but it will be strange to demand from $v_1$ to get the permission of $v_2$ to repel him from the community.

For now we propose the following notion of nodes and edges removal (see protocol \ref{alg: protocol removal}), which will partially allow the community to shrink, as long as it take precaution to remove edges in a valid order. In short, the protocol accepts a request to remove a set of nodes from $A$ if the graph remains highly connected without it. Specifically, it tests whether the graph remains resilient (line 6) and if the bound on corrupt nodes is maintained (line 7). If the conditions do not hold, then it disregards the request to remove the nodes, and do nothing. Removing edges from the underlying graph $G$ is done unconditionally (lines 12-19), as long as the edges are not in $A$.

\begin{algorithm}
\caption{Semi permissioned consensus protocol -- edge removal}\label{alg: protocol removal}
\begin{algorithmic}[1]
\State \ldots
\Function{HandleEvent}{$state$, $event$}
    \State \ldots
    \If{\Call{$\mathit{event}$.getType()}{} is REDUCE\_COMMUNITY}
        \State $A'\gets$\Call{$\mathit{event}$.getCommunity()}{}
        \State $C_1 \gets $\Call{ResilientCommunity}{$A \setminus A'$}
        \State $C_2 \gets $\Call{CorruptionOracle}{$A \setminus A'$}
        \If{$C_1$ and $C_2$}
            \State$A\gets A\setminus A'$
        \EndIf
    \EndIf
    \If{\Call{$\mathit{event}$.getType()}{} is DISCONNECT\_IN\_G}
        \State $\mathit{edges}\gets$\Call{$\mathit{event}$.getEdges()}{}
        \ForAll{$\mathit{edge} \in \mathit{edges}$}
            \If{$\mathit{edge} \notin A$}
                \State $E\gets E\setminus \{\mathit{edge}\}$
            \EndIf
        \EndFor
    \EndIf
    \State \ldots
\EndFunction
\end{algorithmic}
\end{algorithm}

\begin{observation}[Communities can always shrink]
There is always an order of events that will cause a community to shrink all the way to an empty graph.
\end{observation}

The observation holds due to the symmetry between protocol \ref{alg: protocol removal} and protocol \ref{alg: protocol}. Both make sure that in every modification of $A$ the community remains resilient and corruption free before and after the change. In both protocols the addition and removal of edges outside $A$ is without conditions. These symmetries ensure that an `undo' path of events is always feasible. That is, growing the community with a series of growing steps, and then shrinking it back in a reverse order of events will safely shrink the community back to its initial state.

There are two cases where the protocol should allow to safely remove a node from the community. The first is when a node is discovered to be a sybil node. The second is when a genuine node (honest or corrupt) represent an identity that ceased to exist (either passed away or decided to disconnect from the ledger). It is not clear how to always remove a node, when it is situated in a bottleneck that may split the community into two loosely connected parts. We assume, however, that it is always possible to remove any individual node by creating new connections (edges) between the two parts to increase their connectivity.

\section{Joining Communities}\label{section: join}

Up until now new nodes joining the ledger started from scratch, by reading the ledger from another node and updating their internal state to be aligned with the state of the community (see subsection \ref{subsection: joining networks}). In this section we want to consider the more sophisticated scenario where two communities (two sets of nodes), each with its own ledger and history, wish to join together to become a single community. Distributed ledgers are by nature closed environments, as the state of each replica cannot depend on reading of any external inputs, as there are no guarantees that all replicas will read the same input. For this reason, one distributed ledger cannot read the state of another distributed ledger. Nodes participating in one distributed ledger all start from the same initial state, and they all read the same history of events. Two different set of nodes, in two different current state, have no way to reach the same internal state, even if from one point onward they will always accept the same events. More then that, even if one ledger had a magical way to bring itself up to date with the state of the second ledger, there is no way to enforce the second ledger (the second set of nodes) to do the same. The merge must be symmetric. Yet, as each ledger is external environment to the second ledger, there is no methodical way to guarantee symmetric execution.

We therefore look at the merging process as two unidirectional processes. Ledger 1 accepts an event that triggers the assimilation of ledger 2, while in a separate and unrelated event, ledger 2 starts the assimilation of ledger 1. If the result is identical (hash value of the state is the same) then the union of all replicas of both ledgers are in the same state, and are now ready to consume the next event and reach consensus over it over the joined group of replicas. As discussed, the ledgers cannot resolve if the result is identical by themselves and only an external intervention (as human intervention) can arbitrate the question and conclude that the ledgers are indeed merged. Actually, from the ledgers perspective, they are never fully merged, as each ledger will indefinitely maintain the history before the merge which is different then the history of the second ledger. 

The symmetric assimilation of each other's state is implementation dependent. A ledger of currency might need to modify the addresses of all accounts, to avoid conflict between accounts with the same id in both ledgers. A smart contract of voting might need to carefully identify the combined community, while merging duplicates, and then join together all gathered votes between the two communities. We therefore cannot represent a method for this process in the scope of this paper, but we do assume that any distributed ledger application must have a way to correctly merge two partial states into one joined state. The correctness of this merge is not only application dependent, but it might also be preference dependent. One community may prefer to treat duplicate currency account as separate accounts (accidentally having the same id) while another community may prefer to treat duplicate accounts as duplicate accounts of the same agent, and therefore prefer to merge them together into one joined account.

\begin{definition}[\texttt{MergeStates}]
A \emph{\texttt{MergeState}} is an implementation specific external function that receives two different states and returns a single state that is a safe merge of the two states and symmetric, in the following sense: $$\emph{\texttt{MergeState(state1,state2)=MergeState(state2,state1)}}$$
\end{definition}

\begin{algorithm}[p]
\caption{Semi permissioned consensus protocol -- the whole protocol}\label{alg: full protocol}
\begin{algorithmic}[1]
\State init $\mathit{state}$ with an empty graph and $\mathit{logger}$ with an empty log
\Function{HandleEvent}{$\mathit{state}$, $\mathit{event}$}
    \State $G\gets$ \Call{$\mathit{state}$.getGraph()}{}
    \State $(A,V,E)\gets$ \Call{$G$.deconstruct()}{}
    \If{\Call{IsValid}{$\mathit{event}$}}
        \If{\Call{Consensus}{$A$, $\mathit{event}$}}
            \State\Call{$\mathit{logger}$.log}{$\mathit{event}$}
            \If{\Call{$\mathit{event}$.getType()}{} is CONNECT\_IN\_G}
                \State $\mathit{edges}\gets$\Call{$\mathit{event}$.getEdges()}{}
                \ForAll{$\mathit{edge} \in \mathit{edges}$}
                    \State $E\gets E\cup \{\mathit{edge}\}$
                \EndFor
            \EndIf

            \If{\Call{$\mathit{event}$.getType()}{} is DISCONNECT\_IN\_G}
                \State $\mathit{edges}\gets$\Call{$\mathit{event}$.getEdges()}{}
                \ForAll{$\mathit{edge} \in \mathit{edges}$}
                    \If{$\mathit{edge} \notin A$}
                        \State $E\gets E\setminus \{\mathit{edge}\}$
                    \EndIf
                \EndFor
            \EndIf

            \If{\Call{$\mathit{event}$.getType()}{} is EXTEND\_COMMUNITY}
                \State $A'\gets$\Call{$\mathit{event}$.getCommunity()}{}
                \State $\mathit{history}\gets$\Call{$A'$.getLog()}{}
                \State init $\mathit{otherState}$ with an empty state
                \ForAll{$\mathit{event}\in \mathit{history}$}
                    \State\Call{HandleEvent}{$\mathit{otherState}$,$\mathit{event}$}
                \EndFor
                \State $\mathit{joinedState}\gets$\Call{MergeState}{$\mathit{state}$,$\mathit{otherState}$}
                \State $C_1 \gets $\Call{CorruptionOracle}{$A \cup A'$}
                \State $C_2 \gets $\Call{ResilientCommunity}{$A \cup A'$}
                \State $C_3 \gets $\Call{IsValid}{$\mathit{joinedState}$}
                \If{$C_1$ and $C_2$ and $C_3$}
                    \State$A\gets A\cup A'$
                    \State $\mathit{state}\gets\mathit{joinedState}$
                \EndIf
            \EndIf
            \If{\Call{$\mathit{event}$.getType()}{} is REDUCE\_COMMUNITY}
                \State $A'\gets$\Call{$\mathit{event}$.getCommunity()}{}
                \State $C_1 \gets $\Call{ResilientCommunity}{$A \setminus A'$}
                \State $C_2 \gets $\Call{CorruptionOracle}{$A \setminus A'$}
                \If{$C_1$ and $C_2$}
                    \State$A\gets A\setminus A'$
                \EndIf
            \EndIf
        \EndIf
    \EndIf
\EndFunction
\end{algorithmic}
\end{algorithm}

Protocol \ref{alg: full protocol} combines the previous protocols, with the addition that when adding nodes to the community of trusted agents, their history is added to the internal state of the distributed ledger. As explained, this process is unidirectional. When community $A$ reads the history of community $A'$ and adds it to the distributed ledger, it does not merge with the nodes of community $A'$, only create a local copy of their history within the ledger of community $A$. If the same process will simultaneously occur within the network of community $A'$, then the two communities will be aligned (in the same internal state) and will be able to jointly handle any future events. Alternatively, community $A'$ can wait silent until community $A$ finishes to update the joined ledger, and then discard their ($A'$) ledger and join the ledger of $A$ in a similar method to the one presented in subsection \ref{subsection: joining networks}. Successfully merging two ledgers of two communities is therefore an external process that is controlled by the operators of this system and cannot be fully automated. However, Protocol \ref{alg: full protocol} gives the necessary conditions for such an external process to succeed. 

\section{Conclusion}
We showed a safe way to integrate a consensus protocol with a community growth protocol -- based on a trust graph -- to conduct and manage a distributed ledger within a community where all members are taking part in the ledger maintaining process. We showed the conditions under which the safety of the consensus protocol is maintained. For future work it will be of interest to run a stochastic simulation of the above protocol.

\bibliographystyle{plain}
\bibliography{Democracy}

\end{document}